\documentclass{optica-article}

\journal{optcon}

\articletype{Research Article}

\begin{document}

\title{Doubly resonant photonic crystal cavity using merged bound states in the continuum}

\author{Rui Ge,\authormark{1} Xiangmin Liu,\authormark{1} Xiongshuo Yan,\authormark{1} Xianfeng Chen\authormark{1,2} and Yuping Chen\authormark{1*}}

\address{\authormark{1}State Key Laboratory of Advanced Optical Communication Systems and Networks, School of Physics and Astronomy, Shanghai Jiao Tong University, Shanghai 200240, China\\
\authormark{2}Collaborative Innovation Center of Light Manipulations and Applications, Shandong Normal University, Jinan 250358, China\\
}

\email{\authormark{*}ypchen@sjtu.edu.cn} 



\begin{abstract}
In this work, a doubly resonant photonic crystal (PhC) cavity using the merged bound states in the continuum (BICs) is proposed to obtain a higher second harmonic generation (SHG) efficiency. Firstly by scanning geometry parameters the accidental BICs and a band-edge mode outside the light cone can be obtained. Then as the lattice constant or the thickness of the slab is adjusted the accidental BICs will merge. A supercell with large and small holes is constructed and the band-edge mode outside the light cone can be mode-matched with the merged BICs mode. Finally the heterostructure PhC cavity is designed. The merged BICs show a high quality factor for the photonic crystal with finite size. Consequently, the SHG efficiency of the lattice constant near merged BICs of ~6000\% W$^{-1}$ is higher than the one of the isolated BIC.
\end{abstract}

\section{Introduction}
Bound state in the continuum (BIC) has attracted researchers owing to the property of ultra-high quality factor (Q-factor) \cite{zhenbo2014,yuri2021,joseph2021}. BIC is the vortex center in the polarization directions of far-field radiation and can be characterized by the topological charge \cite{zhenbo2014}. BIC can be used to construct laser \cite{huangc2020}, high-quality sensor \cite{wang2022}, opto-mechanical crystal \cite{liusy2022}, and chiral-emission metasurface \cite{zhangxd2022}. By splitting the BICs up, the unidirectional guided resonance can be achieved \cite{yinxf2020,lee2020,zhangzi2021,zyx2021}. Utilizing the concept of BIC, the doubly resonant photonic crystal (PhC) cavity can be designed and it shows the improved nonlinear frequency conversion process \cite{minkov2019,minkov2020,minkov2021}. In doubly resonant cavity, a BIC mode is mode-matched with a band-edge mode outside the light cone and the generated second-harmonic mode can be collected within a small angle. However, the product of Q-factors at the band-edge mode and the BIC mode still show room to improve. \par 

Recently, researchers start to notice a special kind of BIC called "merged-BICs" \cite{jinji2019}, which is also defined as the "super-BIC" \cite{Hwang2021}. The merged BICs is formed by merging multiple BIC modes into one point. Traditional BIC-based device usually suffers from scattering loss which is caused by the coupling with nearby radiative modes. The strategy of merging BICs into one point can enhance the Q-factors of nearby resonances in the same band, which will result in the BIC being robust against inevitable fabrication imperfection \cite{jinji2019,kangmeng2019,kangmeng2022,zhaochen2022}. In addition, researchers subsequently find other merits of the merged BICs resulting from this trait. For example, the radiative Q-factor at the merged BICs point is generally much larger than the one at pre-merging point or isolated BIC point for a device with finite size \cite{Hwang2021}. Merged BICs have already been widely applied to ultra-low threshold laser \cite{Hwang2021}, chiral resonator \cite{wan2022}, and acoustic resonator \cite{huang2022}. Moreover, these devices show improved quality in comparison to the device that uses isolated BIC mode. \par

Owing to the large Q-factor of the merged BICs in a PhC with finite size, the doubly resonant PhC cavity based on that may show improved nonlinear conversion efficiency. However, it requires that the resonant mode at the second-harmonic frequency must be the merged BICs mode, and a band-edge mode at the fundamental frequency must be mode-matched with that. Simultaneously meeting these conditions is inconvenient. Meanwhile, nonlinear conversion efficiency also depends on the Q-factor at fundamental frequency and the nonlinear overlapping factor, which should also be considered. In this paper, we take lithium niobate (LN) PhC as an example to demonstrate that utilizing the supercell constructed by the large and small air holes can easily achieve these goals. The BIC-based LN photonic devices are already theoretically and experimentally exhibited \cite{kanglei2021,yefan2022,zhang2022,huangzhi2022,zhengze2022} and are proven to be an ideal platform for nonlinear process. In previous works, our group has proposed the beam splitter \cite{duan2016}, nonlinear cavity \cite{jiang2018}, logic gate \cite{lu2019}, and valley waveguide \cite{ge2021} based on lithium niobate PhC. Firstly the band property of the proposed PhC supercell is analyzed and the approach for matching a band-edge mode outside the light cone with the merged BICs is introduced. Then a heterostructure PhC cavity is considered. After determining the geometry parameters the device can meet the mode-matching condition. Finally, we estimated the SHG efficiency of the device at different lattice constants.

\section{Model and theory}

Before discussing our simulations we must firstly review the four requirements to design a doubly resonant PhC cavity using an isolated BIC \cite{minkov2019}: (1) The fundamental mode must be a band-edge mode outside the light cone. (2) The second-harmonic mode must be a BIC mode at the $\Gamma$ point. (3) Those modes are either a maximum or a minimum in their bands. (4) The periodic electric field of those modes has a nonzero nonlinear overlapping factor. Now an additional requirement must be satisfied: the BIC mode must be the merged BICs mode. \par

\begin{figure}[!ht]
\centering
\includegraphics[width=0.55\linewidth]{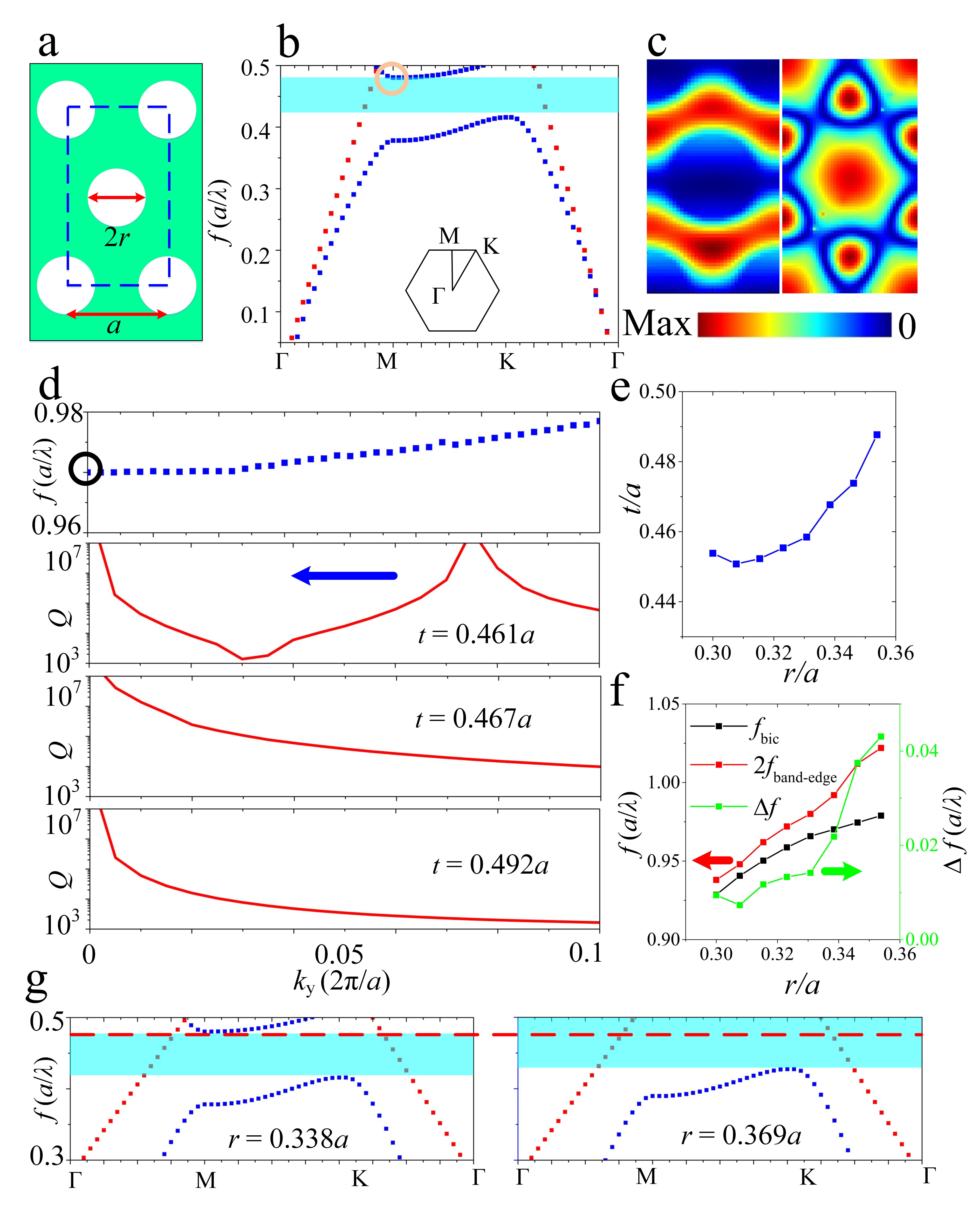}
\caption{(a) Schematic of the simulated unit cell. (b) Band diagram at the fundamental frequency. (c) Field profiles of the modes inside the circles at fundamental and second-harmonic frequencies. (d) Band diagram at second-harmonic frequency together with the Q-factor variations of the band at different thicknesses. (e) Variations of the thickness where the merged BICs locates versus the radii of holes. (f) Dependence of radius of the holes on the doubled frequency of the band-edge mode outside the light cone, the frequency of the merged BICs, and their difference, respectively. (g) Band diagram at the fundamental frequency with $r$=0.338$a$ and $r$=0.369$a$.}
\label{fig1}
\end{figure}

Now back to our design, the first step is to find an accidental BIC mode that satisfies $f_{acc-BIC}$$\approx2$$f_{band-edge}$, where $f_{acc-BIC}$ and $f_{band-edge}$ represent the frequencies of accidental BIC and band edge mode. It is generally simple to obtain an accidental BIC \cite{zhenbo2014}, while the latter requirement is, to some extent, difficult to meet. Here we choose the structure in Fig. \ref{fig1}(a) which is infinitely extended. The air holes are etched in the suspended slab and are arranged in hexagonal lattice. The material of the slab is chosen as LN, which has outstanding nonlinear properties \cite{huangzhi2022}. Current etching technique for LN has demonstrated 85 degrees sidewall angle of holes \cite{liang2017,limx2019,lim2019}. The dispersion of the material should be considered \cite{minkov2019}. The optical axis of LN is set to $z$ direction and consequently, the nonlinear tensor $d_{31}$ is responsible for the second-harmonic generation process. Moreover, we mainly focus on the transverse electric (TE) band at the fundamental frequency and the transverse magnetic (TM) band at second-harmonic frequency. Intuitively once the $d_{33}$ tensor is used the nonlinear conversion efficiency will be higher. However, the merged BICs mode that satisfies four requirements is hard to obtain. Detailed simulation results are discussed in the Appendix part. The refractive index tensors are $n_{x}$=2.2111, $n_{y}$=2.2111, and $n_{z}$=2.1376 near 1550 nm and $n_{x}$=2.2587, $n_{y}$=2.2587, and $n_{z}$=2.1784 near 775 nm \cite{saravi2015}. After numerous simulations with parameter sweep, the lattice constant is set to $a$=650 nm and the radius of holes is set to $r$=0.338$a$=220 nm. The thickness of the slab is set to $t$=0.461$a$=300 nm. The dashed box in Fig. \ref{fig1}(a) indicates the simulation area. All the simulations in our work are completed using three-dimensional finite difference time domain (3D-FDTD) method. \par

The TE band diagram of the structure at the fundamental frequency is plotted in Fig. \ref{fig1}(b). The blue dots indicate the bulk bands while the red dots indicate the light line. The band gap region is marked by a blue rectangle and the Brillouin zone used for the sweep is also plotted. According to Ref.\cite{minkov2019}, the band edge mode at upper bulk band near the band gap can be selected for the frequency conversion as shown in Fig. \ref{fig1}(b). The $|H_{z}|$ field of the chosen band-edge mode outside the light cone inside the brown circle is demonstrated in the left-half part in Fig. \ref{fig1}(c). The absolute value is shown to search the position with the maximum energy. The TM band diagram of the structure at the second-harmonic frequency is plotted in the top part in Fig. \ref{fig1}(d). The $|E_{z}|$ field of the chosen BIC mode inside the black circle is demonstrated in the right-half part of Fig. \ref{fig1}(c). From the field profiles it can be inferred that the energy in chosen TE mode is mainly located in two side lobes around the hole, while the the energy in chosen TM mode is mainly located in six clusters around the hole. The TM band indicates a symmetry-protected BIC and an accidental BIC, which is verified by calculating the Q-factors of the band. The bottom part of Fig. \ref{fig1}(d) shows that at $k_{y}$=0 and $k_{y}$=0.075 ($2\pi/a$) the Q-factors become infinite when $t=0.461a$. Previous work has already demonstrated that the FDTD method can simulate above ~10$^{6}$ Q-factor \cite{minkov2019}. As for the TE band, the Q-factor at $\Gamma$ point is infinite owing to that it locates outside the light cone. According to the conservation law of the topological charge, the accidental BIC mode will move its position in $k$ space as the thickness or the lattice constant of PhC gradually changes while the symmetry-protected BIC will be fixed \cite{zhenbo2014,kangmeng2022}. Here the thickness of the slab is gradually increased, and it can be observed that the accidental BIC mode gradually merges toward $\Gamma$ point and finally the merged BICs mode can be obtained at $t=0.467a$. The blue arrow indicates the moving direction as shown in Fig. \ref{fig1}(d). At the merged BICs point the Q-factor decreases extremely slowly as $k_{y}$ increases, which is the evidence of the merged BICs. As the thickness becomes further larger, the topological charges will cancel each other and the mode at $\Gamma$ point will be a symmetry-protected BIC again instead of the merged BICs. Although the structure in Fig. \ref{fig1}(a) does not work near 1550 nm, according to the scaling rule of PhC by changing the absolute value of the lattice constant the device can work near 1550 nm. \par

Next, the requirement of the mode-matching condition must be satisfied, i.e., $f_{acc-BIC}$=2$f_{band-edge}$ must be satisfied. In the current geometry parameter, the mode-mismatch $\Delta$$f$=$f_{acc-BIC}$-2$f_{band-edge}$ still cannot attain zero. According to Ref. \cite{minkov2019}, the mode-matching can be achieved by adjusting one geometry parameter like $r$ or $t$ as the $f_{acc-BIC}$ and 2$f_{band-edge}$ change with different speeds versus $r$ or $t$. However, once $r$ or $t$ is adjusted, the BIC will no longer be the merged state. Intuitively, both $r$ and $t$ can be varied for the mode-matching. Here the $r$ is gradually varied, and according to the conservation law of the topological charge for each $r$ there must be a certain $t$ where the merged BICs locates. In Fig. \ref{fig1}(e) we show the thickness of the slab where the merged BICs locates versus numerous radii of holes and the results. Interestingly, as the radius decreases the thickness attains a saturation value. Variations of $f_{acc-BIC}$, 2$f_{band-edge}$ and $\Delta$$f$ versus $r$ are plotted in Fig. \ref{fig1}(f) and it should be noted that for each different $r$ in the abscissa of Fig. \ref{fig1}(f) its corresponding $t$ is also different to ensure each point is a state where BICs merge. As $r$ decreases the $\Delta$$f$ also attains a saturation value and cannot approach zero. The simulated results indicate that adjusting both $r$ and $t$ is not sufficient to realize the mode-matching condition. \par

In Fig. \ref{fig1}(g) the band diagram at the fundamental frequency for $r$=0.338$a$ and $r$=0.369$a$ are plotted. It can be seen that the whole bands have moved into the higher frequency when $r$=0.369$a$. That's to say, the selected band with $r$=0.338$a$ locates at the band gap region in the diagram with $r$=0.369$a$. Consequently, the PhC with larger $r$ can be used to confine light at fundamental frequency in our device. \par

\begin{figure}[!ht]
\centering
\includegraphics[width=0.55\linewidth]{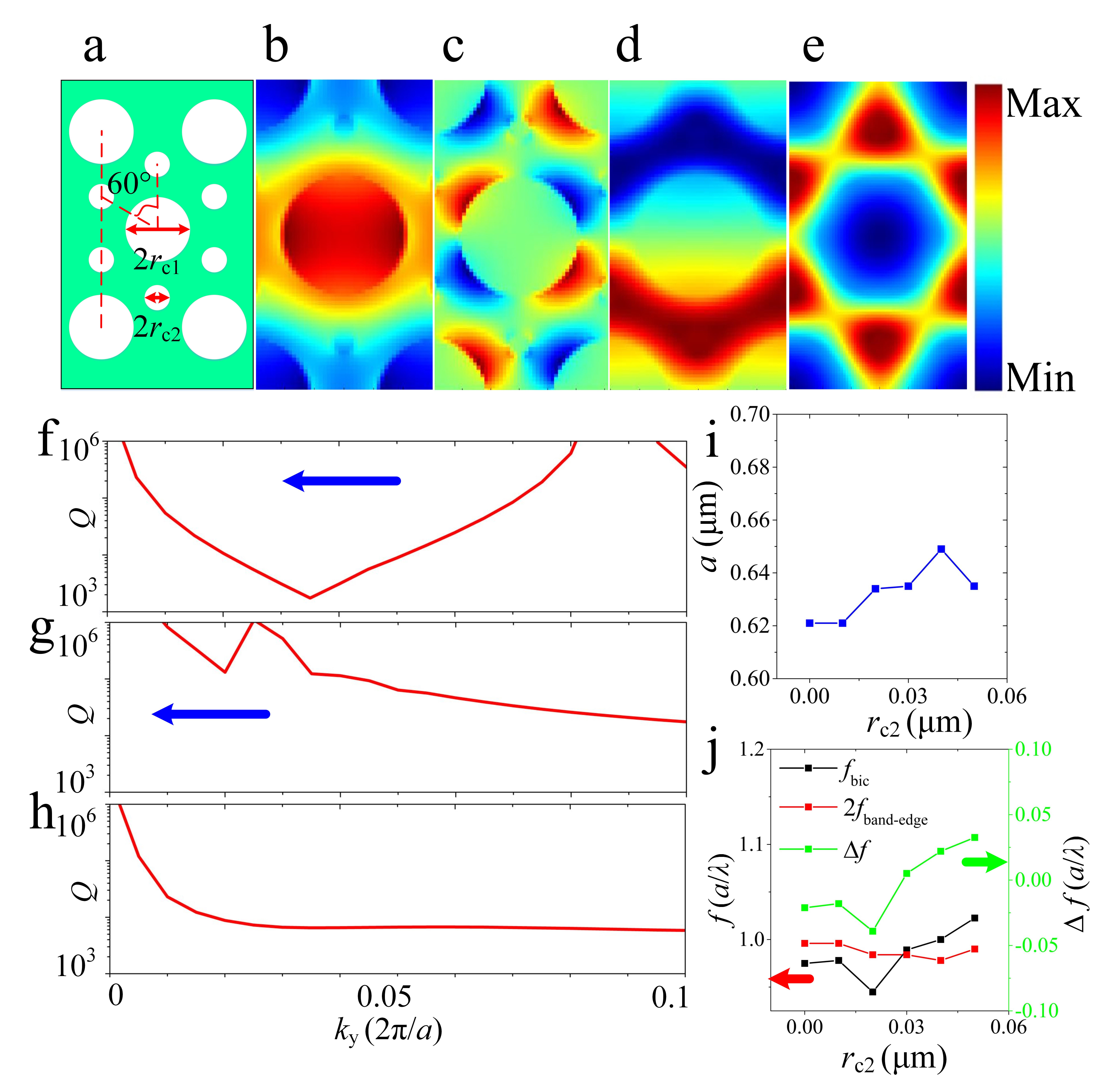}
\caption{(a) Schematic of the simulated unit cell. (b) $E_{x}$, (c) $E_{y}$, and (d) $H_{z}$ field profiles of the modes at the fundamental frequency, and (e) $E_{z}$ field profile at the second-harmonic frequency. (f)-(g) Q-factor variations of the band with the lattice constant of 653 nm, 649 nm and 640 nm, respectively. (i) Variations of the lattice constant where the merged BICs locates versus radii of holes. (j) Dependence of radius of holes on the doubled frequency of the band-edge mode outside the light cone, the frequency of the merged BICs mode, and their difference, respectively.}
\label{fig2}
\end{figure}

To make the band-edge mode outside the light cone mode match the merged BICs mode, six new small holes around the large hole are added as shown in Fig. \ref{fig2}(a). The implementation does not break the C$_{6v}$ symmetry of the system and the merged topological charge at $\Gamma$ point will not suddenly disappear \cite{Yoda2020}. Constructing the supercell with the large and the small holes has already been used to achieve BIC-based negative refraction \cite{lari2021}. Here the parameters are $a$=650 nm, $t$=286 nm, and $r_{c1}$=210 nm, respectively. $r_{c1}$ and $r_{c2}$ represent the radius of large holes and small holes. The mode profiles of the merged BICs at $\Gamma$ point are shown in Figs. \ref{fig2}(b)-\ref{fig2}(d). The mode profile of the band-edge mode outside the light cone is shown in Fig. \ref{fig2}(e). Here the profiles are plotted to ensure the fundamental and the second-harmonic modes have a nonzero nonlinear overlapping factor. The principle of this design mainly lies in the different field distributions of the fundamental and the second-harmonic modes. We expected that adding small holes influences the wavelength of BIC mode owing to that the energy mainly distributes over the area where the small holes are constructed. As the $a$ is gradually adjusted, the accidental BICs will merge at $\Gamma$ point and the merged BICs can be obtained. When $r_{c2}$=50 nm, the Q-factor variations of the band at the lattice constant of 653 nm, 649 nm, and 640 nm versus $k$ are plotted as shown in Figs. \ref{fig2}(f)-\ref{fig2}(h), respectively. The blue arrow indicates the moving direction of the accidental BIC. In comparison to the results in Fig. \ref{fig1}(d) the simulated maximum Q-factor becomes lower. The reason lies in that adding small structure enhances the difficulties of dividing the mesh. It reminds us that for each lattice constant, we will select a suitable $r_{c2}$ to achieve the mode-matching condition. In Fig. \ref{fig2}(i) the lattice constant variations of the slab where the merged BICs locates for numerous radii of small holes are shown. For each merged BICs point $f_{acc-BIC}$, 2$f_{band-edge}$ and $\Delta$$f$ versus $r$ are shown in Fig. \ref{fig2}(j) and similarly it should be noted that for each different $r_{c2}$ in the abscissa its corresponding $t$ is also different to ensure each point is a merged BICs point. As we predicted the 2$f_{band-edge}$ nearly stays steady as $r_{c2}$ varies while the $f_{acc-BIC}$ changes fiercely. $\Delta$$f$ finally crosses the zero point and the matching condition can be realized. The method of adding small holes can adjust the $\Delta$$f$ over a large range.

\begin{figure}[!ht]
\centering
\includegraphics[width=0.55\linewidth]{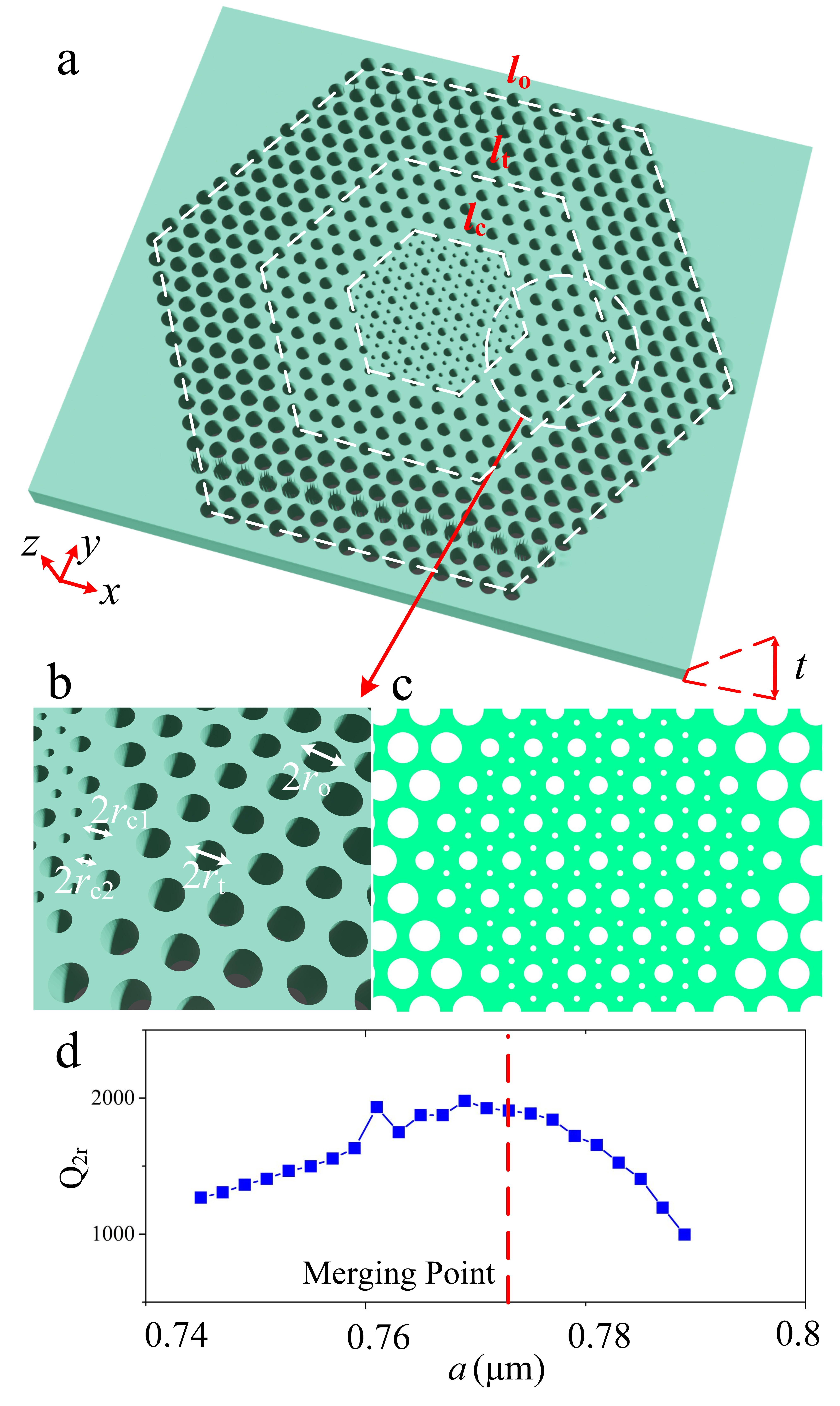}
\caption{(a) Model of simulated heterostructure PhC cavities. (b) Three dimensional and (c) two dimensional enlarged views of the proposed device. (d) Dependence of radiative Q-factors of the PhC cavity versus the lattice constants.}
\label{fig3}

\end{figure}
\begin{figure}[!ht]
\centering
\includegraphics[width=0.55\linewidth]{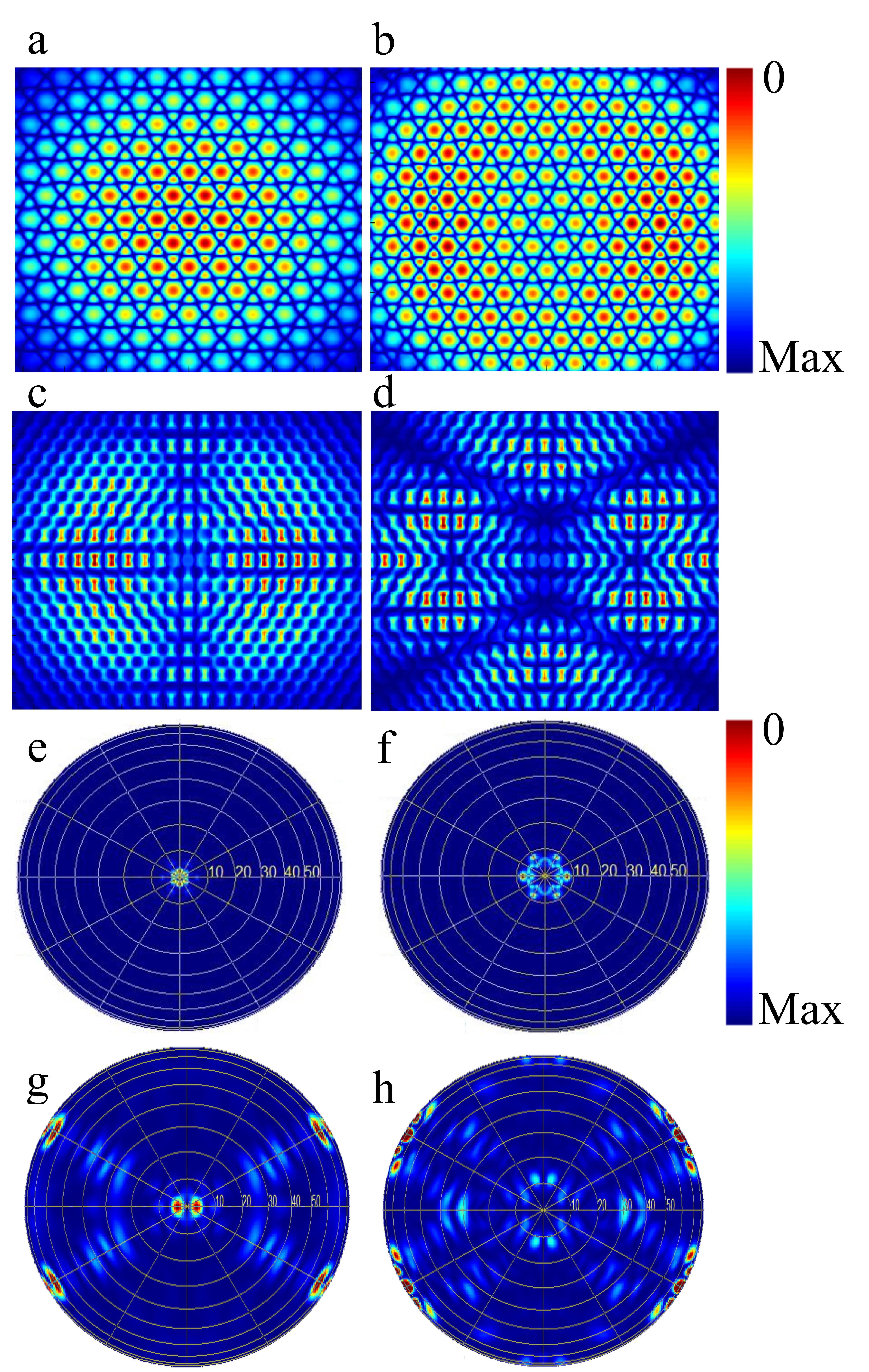}
\caption{Field profiles of the modes at (a) 768.6 nm, (b) 763.6 nm, (c) 1541.6 nm, and (d) 1533.6 nm, respectively. Polar far-field emission profiles of the modes at (e) 768.6 nm, (f) 763.6 nm, (g) 1541.6 nm, and (h) 1533.6 nm, respectively.}
\label{fig4}
\end{figure}

\begin{figure}[!ht]
\centering
\includegraphics[width=0.7\linewidth]{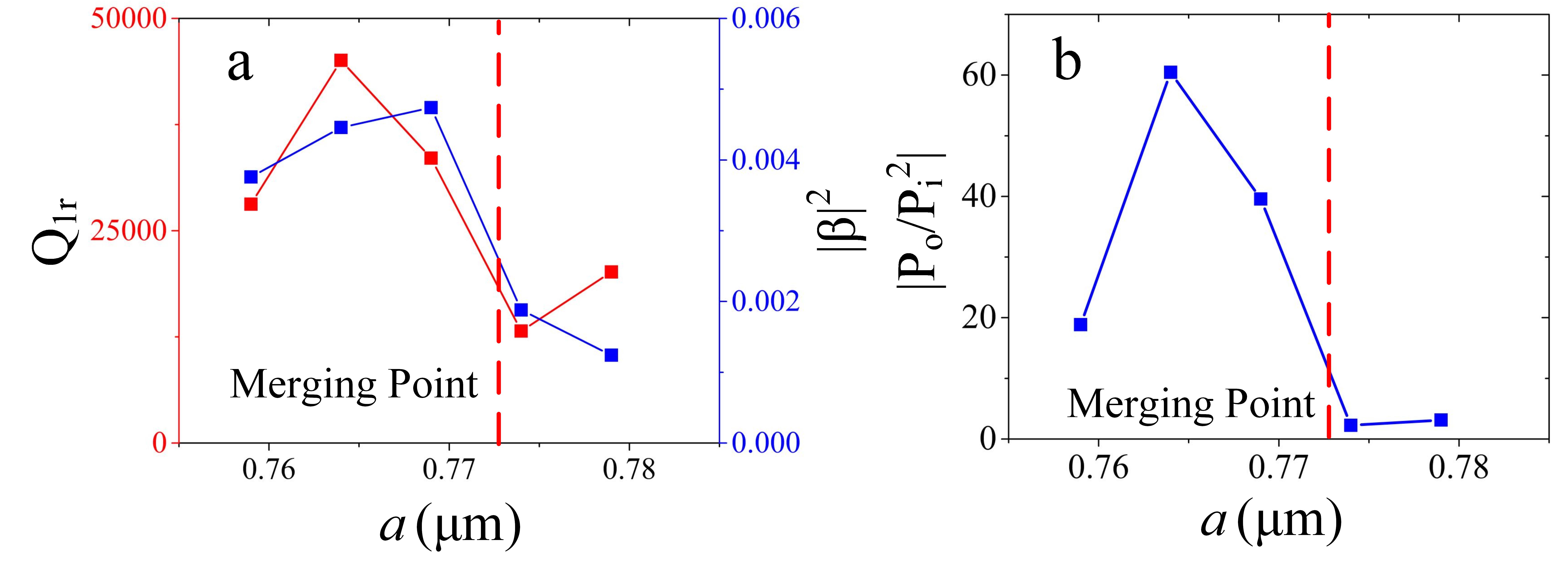}
\caption{(a) Q-factors at the fundamental frequencies versus lattice constants. (b) The square of nonlinear overlapping factors versus lattice constants. (c) Second harmonic generation efficiency versus lattice constants.}
\label{fig5}
\end{figure}

\section{Results and discussions}
In this section, the heterostructure PhC cavity is considered as shown in Fig. \ref{fig3}(a). The lattice constant of the device is enlarged compared with that in Fig. \ref{fig1} to ensure the cavity works near 1550 nm. The small and the large holes possess the same lattice constant. The optical axis of the material points to $z$ direction. Like Ref. \cite{minkov2019} the heterostructure PhC can be divided into core region, transition region, and outer region. These regions possess the same lattice constant. The outer region is constructed for confining the photons near 1550 nm and the transition region is just for improving the Q-factor \cite{minkov2019}. The radii of holes in core, transition, and outer regions are $r_{c1}$=254 nm, $r_{t}$=260 nm, and $r_{o}$=270 nm. By carefully adjusting the geometry parameters the mode-matching condition can be satisfied. The side lengths of three regions are $l_{c}$=10$a$, $l_{t}$=14$a$, and $l_{o}$=24$a$, respectively. Only in core region there are small holes and $r_{c2}$=40 nm. The thickness is $t$=345 nm. When the lattice constant is $a$=769 nm, the device supports the merged BICs mode at 768.6 nm and a band-edge mode at 1541.1 nm. \par

In Fig. \ref{fig3}(d) the dependence of radiative Q-factor of the BIC mode on the lattice constant of the structure is plotted. When $a$=769 nm, the Q-factor obtains its maximum value of 1979, while for the isolated BIC it is below 1000. This is the evidence of the merged BICs mode for a finite-size PhC \cite{Hwang2021}. It should be noted that according to Ref. \cite{Hwang2021} the $a$ with the maximum Q-factor usually differs from the $a$ of the merged BICs point as shown in Fig. \ref{fig3}(d). The $a$ in our infinite system shows the merged BICs point at $a$=773 nm. \par

To characterize the performance of the proposed device, the field profiles and the far-field emission profiles of the merged BICs mode and the band-edge mode outside the light cone are plotted. The $|E_{z}|$ field profile of the merged BICs mode at 768.6 nm is shown in Fig. \ref{fig4}(a). The field profile of the higher-order mode at 763.6 nm is also shown in Fig. \ref{fig4}(b). In  Figs. \ref{fig4}(c)-\ref{fig4}(d) the $|H_{z}|$ field of the fundamental and the higher-order band-edge mode are plotted and they locate at 1541.1 nm and 1533.6 nm. In Figs. \ref{fig4}(e)-\ref{fig4}(f) the polar far-field emission profiles of the merged BICs mode in the upper half space are obtained. As we predicted the far-field emission is the hollow beam, which is determined by the vortex nature of the BIC \cite{minkov2019}. In Figs. \ref{fig4}(g)-\ref{fig4}(h) the far-field emission profiles of the band-edge mode and the higher-order band-edge mode in the upper half space are obtained. The results indicate that the far-field emissions at 768.6 nm and 1541.1 nm are highly collimated around the normal incidence, which leads to efficient excitation and collection of nonlinear signal \cite{minkov2019}. \par

At last we estimate the performance of the proposed device as a nonlinear cavity. The second-harmonic generation efficiency can be calculated using the formula \cite{minkov2019,minkov2021}:

\begin{equation}
\frac{P_{o}}{P_{i}^{2}}=\frac{8}{\omega_{1}}(\frac{\chi^{(2)}}{\sqrt{\epsilon_{0}\lambda_{FH}^{3})}})^{2}|\overline{\beta}|^{2}Q_{FH}^{2}Q_{SH}
\label{eq:1}
\end{equation}

Where $\epsilon_{0}$ is the vacuum permittivity. $\overline{\beta}$ can be determined by \cite{minkov2019,minkov2021,chenya2021}:

\begin{equation}
\overline{\beta}=\frac{\int d\textbf{r}\sum_{ijk}\overline{\chi}_{ijk}E_{2\omega i}^{*}E_{\omega j}E_{\omega k}}{(\int d\textbf{r}\epsilon_{\omega}(\textbf{r})|\textbf{E}_{\omega}|^{2})(\int d\textbf{r}\epsilon_{2\omega}(\textbf{r})|\textbf{E}_{2\omega}|^{2})^{1/2}}\lambda_{FH}^{3/2}
\label{eq:2}
\end{equation}

Where $\lambda_{FH}$ is the wavelength of the band-edge mode, $\overline{\chi_{ijk}}$ is the dimensionless nonlinear tensor elements. Here we assumed that the extrinsic Q-factor is infinite and the perfect pumping and collecting condition can be satisfied. We noted that most energy is distributed in the center of the cavity and electric field data in the boundary of the cavity is not considered for convenience of calculation. Consequently the $|\overline{\beta}|$ is a little larger than its actual value but it does not influence its relative value for different lattice constants. \par

It can be inferred that the theoretical second-harmonic generation efficiency is mainly determined by the Q-factors at the fundamental and the second harmonic frequencies and the nonlinear overlapping factor. In our work, the PhC slab is constructed by LN, and the optical axis is pointed to a fixed direction. The resonant frequencies of modes at different constants near 769 nm are similar. In Fig. \ref{fig5}(a)-\ref{fig5}(b) we demonstrate the Q-factors of the band-edge mode outside the light cone and the square of nonlinear overlapping factors versus lattice constants. The results indicate that they also show a peak near the merged BICs point. Consequently, the second-harmonic generation efficiency will obtain its maximum value near the merged BICs point as shown in Fig. \ref{fig5}(c). The maximum value is ~60 W$^{-1}$ (6000\% W$^{-1}$). The value is far large than that of isolated BIC.

\section{Conclusion}

In conclusion, we proposed a doubly resonant LN photonic crystal cavity using the merged BICs to achieve a higher SHG efficiency. The unit cell of large and small holes is established and a band-edge mode at the fundamental frequency will be mode-matched with the merged BICs mode. It can be found that the SHG of the merged BICs of ~60 W$^{-1}$ (6000\% W$^{-1}$) is higher than the one of the isolated BIC. Our design recipe is not limited to LN and can be extended to other nonlinear materials like GaN or AlGaAs. In addition, except for the second-harmonic generation, our design can apply to the parametric down-conversion process. Except for the higher Q-factor at the merged BIC point for a finite-size PhC, the merit of the merged BICs also includes the robustness of Q-factor against the random fluctuations on radii or lattice constants of the holes. Consequently, the nonlinear conversion efficiency may shows the slighter degradation compared with the one of isolated BIC against disorder. However, it is just a hypothesis and requires our future simulation to verify. This work is expected to broader the application of the merged BICs in nonlinear photonic area.

\par

  This work was supported by the National Natural Science Foundation of China (Grant Nos. 91950107, and 12134009), the National Key R\&D Program of China (Grant Nos. 2019YFB2203501), Shanghai Municipal Science and Technology Major Project (2019SHZDZX01-ZX06), and SJTU No. 21X010200828.


\bibliography{sample}

\begin{thebibliography}{10}
\newcommand{\enquote}[1]{``#1''}

\bibitem{zhenbo2014}
B.~Zhen, C.~W. Hsu, L.~Lu, A.~D. Stone, and M.~Solja{\v{c}}i{\'c},
  \enquote{Topological nature of optical bound states in the continuum,}
  {\protect\JournalTitle{Physical Review Letters}} \textbf{113}, 257401 (2014).

\bibitem{yuri2021}
W.~Liu, W.~Liu, L.~Shi, and Y.~Kivshar, \enquote{Topological polarization
  singularities in metaphotonics,} {\protect\JournalTitle{Nanophotonics}}
  \textbf{10}, 1469--1486 (2021).

\bibitem{joseph2021}
S.~Joseph, S.~Pandey, S.~Sarkar, and J.~Joseph, \enquote{Bound states in the
  continuum in resonant nanostructures: an overview of engineered materials for
  tailored applications,} {\protect\JournalTitle{Nanophotonics}}  (2021).

\bibitem{huangc2020}
C.~Huang, C.~Zhang, S.~Xiao, Y.~Wang, Y.~Fan, Y.~Liu, N.~Zhang, G.~Qu, H.~Ji,
  J.~Han \emph{et~al.}, \enquote{Ultrafast control of vortex microlasers,}
  {\protect\JournalTitle{Science}} \textbf{367}, 1018--1021 (2020).

\bibitem{wang2022}
X.~Wang, J.~Xin, Q.~Ren, H.~Cai, J.~Han, C.~Tian, P.~Zhang, L.~Jiang, Z.~Lan,
  J.~You \emph{et~al.}, \enquote{Plasmon hybridization induced by quasi bound
  state in the continuum of graphene metasurfaces oriented for high-accuracy
  polarization-insensitive two-dimensional sensors,}
  {\protect\JournalTitle{Chinese Optics Letters}} \textbf{20}, 042201 (2022).

\bibitem{liusy2022}
S.~Liu, H.~Tong, and K.~Fang, \enquote{Optomechanical crystal with bound states
  in the continuum,} {\protect\JournalTitle{Nature Communications}}
  \textbf{13}, 1--7 (2022).

\bibitem{zhangxd2022}
X.~Zhang, Y.~Liu, J.~Han, Y.~Kivshar, and Q.~Song, \enquote{Chiral emission
  from resonant metasurfaces,} {\protect\JournalTitle{Science}} \textbf{377},
  1215--1218 (2022).

\bibitem{yinxf2020}
X.~Yin, J.~Jin, M.~Solja{\v{c}}i{\'c}, C.~Peng, and B.~Zhen,
  \enquote{Observation of topologically enabled unidirectional guided
  resonances,} {\protect\JournalTitle{Nature}} \textbf{580}, 467--471 (2020).

\bibitem{lee2020}
S.-G. Lee, S.-H. Kim, and C.-S. Kee, \enquote{Bound states in the continuum
  (bic) accompanied by avoided crossings in leaky-mode photonic lattices,}
  {\protect\JournalTitle{Nanophotonics}} \textbf{9}, 4373--4380 (2020).

\bibitem{zhangzi2021}
Z.~Zhang, X.~Yin, Z.~Chen, F.~Wang, W.~Hu, and C.~Peng, \enquote{Observation of
  intensity flattened phase shifting enabled by unidirectional guided
  resonance,} {\protect\JournalTitle{Nanophotonics}} \textbf{10}, 4467--4475
  (2021).

\bibitem{zyx2021}
Z.~Zhang, X.~Yin, Z.~Chen, F.~Wang, W.~Hu, and C.~Peng, \enquote{Observation of
  intensity flattened phase shifting enabled by unidirectional guided
  resonance,} {\protect\JournalTitle{Nanophotonics}} \textbf{10}, 4467--4475
  (2021).

\bibitem{minkov2019}
M.~Minkov, D.~Gerace, and S.~Fan, \enquote{Doubly resonant $\chi$ (2) nonlinear
  photonic crystal cavity based on a bound state in the continuum,}
  {\protect\JournalTitle{Optica}} \textbf{6}, 1039--1045 (2019).

\bibitem{minkov2020}
J.~Wang, M.~Clementi, M.~Minkov, A.~Barone, J.-F. Carlin, N.~Grandjean,
  D.~Gerace, S.~Fan, M.~Galli, and R.~Houdr{\'e}, \enquote{Doubly resonant
  second-harmonic generation of a vortex beam from a bound state in the
  continuum,} {\protect\JournalTitle{Optica}} \textbf{7}, 1126--1132 (2020).

\bibitem{minkov2021}
S.~Zanotti, M.~Minkov, S.~Fan, L.~C. Andreani, and D.~Gerace,
  \enquote{Doubly-resonant photonic crystal cavities for efficient
  second-harmonic generation in iii--v semiconductors,}
  {\protect\JournalTitle{Nanomaterials}} \textbf{11}, 605 (2021).

\bibitem{jinji2019}
J.~Jin, X.~Yin, L.~Ni, M.~Solja{\v{c}}i{\'c}, B.~Zhen, and C.~Peng,
  \enquote{Topologically enabled ultrahigh-q guided resonances robust to
  out-of-plane scattering,} {\protect\JournalTitle{Nature}} \textbf{574},
  501--504 (2019).

\bibitem{Hwang2021}
M.-S. Hwang, H.-C. Lee, K.-H. Kim, K.-Y. Jeong, S.-H. Kwon, K.~Koshelev,
  Y.~Kivshar, and H.-G. Park, \enquote{Ultralow-threshold laser using
  super-bound states in the continuum,} {\protect\JournalTitle{Nature
  Communications}} \textbf{12}, 1--9 (2021).

\bibitem{kangmeng2019}
M.~Kang, S.~Zhang, M.~Xiao, and H.~Xu, \enquote{Merging bound states in the
  continuum at off-high symmetry points,} {\protect\JournalTitle{Physical
  Review Letters}} \textbf{126}, 117402 (2021).

\bibitem{kangmeng2022}
M.~Kang, L.~Mao, S.~Zhang, M.~Xiao, H.~Xu, and C.~T. Chan, \enquote{Merging
  bound states in the continuum by harnessing higher-order topological
  charges,} {\protect\JournalTitle{Light: Science \& Applications}}
  \textbf{11}, 1--9 (2022).

\bibitem{zhaochen2022}
C.~Zhao, W.~Chen, J.~Wei, W.~Deng, Y.~Yan, Y.~Zhang, and C.-W. Qiu,
  \enquote{Electrically tunable and robust bound states in the continuum
  enabled by 2d transition metal dichalcogenide,}
  {\protect\JournalTitle{Advanced Optical Materials}} p. 2201634 (2022).

\bibitem{wan2022}
S.~WAN, K.~WANG, F.~WANG, C.~GUAN, W.~LI, J.~LIU, A.~BOGDANOV, P.~A. BELOV, and
  J.~SHI, \enquote{Topologically enabled ultrahigh-q chiroptical resonances by
  merging bound states in the continuum,} {\protect\JournalTitle{Optics
  Letters}} \textbf{47}, 3291--3294 (2022).

\bibitem{huang2022}
L.~Huang, B.~Jia, Y.~K. Chiang, S.~Huang, C.~Shen, F.~Deng, T.~Yang, D.~A.
  Powell, Y.~Li, and A.~E. Miroshnichenko, \enquote{Topological supercavity
  resonances in the finite system,} {\protect\JournalTitle{Advanced Science}}
  p. 2200257 (2022).

\bibitem{kanglei2021}
L.~Kang, H.~Bao, and D.~H. Werner, \enquote{Efficient second-harmonic
  generation in high q-factor asymmetric lithium niobate metasurfaces,}
  {\protect\JournalTitle{Optics Letters}} \textbf{46}, 633--636 (2021).

\bibitem{yefan2022}
F.~Ye, Y.~Yu, X.~Xi, and X.~Sun, \enquote{Second-harmonic generation in
  etchless lithium niobate nanophotonic waveguides with bound states in the
  continuum,} {\protect\JournalTitle{Laser \& Photonics Reviews}} \textbf{16},
  2100429 (2022).

\bibitem{zhang2022}
X.~Zhang, L.~He, X.~Gan, X.~Huang, Y.~Du, Z.~Zhai, Z.~Li, Y.~Zheng, X.~Chen,
  Y.~Cai \emph{et~al.}, \enquote{Quasi-bound states in the continuum enhanced
  second-harmonic generation in thin-film lithium niobate,}
  {\protect\JournalTitle{Laser \& Photonics Reviews}} \textbf{16}, 2200031
  (2022).

\bibitem{huangzhi2022}
Z.~Huang, K.~Luo, Z.~Feng, Z.~Zhang, Y.~Li, W.~Qiu, H.~Guan, Y.~Xu, X.~Li, and
  H.~Lu, \enquote{Resonant enhancement of second harmonic generation in
  etchless thin film lithium niobate heteronanostructure,}
  {\protect\JournalTitle{Science China Physics, Mechanics \& Astronomy}}
  \textbf{65}, 1--8 (2022).

\bibitem{zhengze2022}
Z.~Zheng, L.~Xu, L.~Huang, D.~Smirnova, P.~Hong, C.~Ying, and M.~Rahmani,
  \enquote{Boosting second-harmonic generation in the linbo 3 metasurface using
  high-q guided resonances and bound states in the continuum,}
  {\protect\JournalTitle{Physical Review B}} \textbf{106}, 125411 (2022).

\bibitem{duan2016}
S.~Duan, Y.~Chen, G.~Li, C.~Zhu, and X.~Chen, \enquote{Broadband polarization
  beam splitter based on a negative refractive lithium niobate photonic crystal
  slab,} {\protect\JournalTitle{Chinese Optics Letters}} \textbf{14}, 042301
  (2016).

\bibitem{jiang2018}
H.~Jiang, H.~Liang, R.~Luo, X.~Chen, Y.~Chen, and Q.~Lin, \enquote{Nonlinear
  frequency conversion in one dimensional lithium niobate photonic crystal
  nanocavities,} {\protect\JournalTitle{Applied Physics Letters}} \textbf{113},
  021104 (2018).

\bibitem{lu2019}
C.~Lu, B.~Zhu, C.~Zhu, L.~Ge, Y.~Liu, Y.~Chen, and X.~Chen,
  \enquote{All-optical logic gates and a half-adder based on lithium niobate
  photonic crystal micro-cavities,} {\protect\JournalTitle{Chinese Optics
  Letters}} \textbf{17}, 072301 (2019).

\bibitem{ge2021}
R.~Ge, X.~Yan, Y.~Chen, and X.~Chen, \enquote{Broadband and lossless lithium
  niobate valley photonic crystal waveguide,} {\protect\JournalTitle{Chinese
  Optics Letters}} \textbf{19}, 060014 (2021).

\bibitem{liang2017}
H.~Liang, R.~Luo, Y.~He, H.~Jiang, and Q.~Lin, \enquote{High-quality lithium
  niobate photonic crystal nanocavities,} {\protect\JournalTitle{Optica}}
  \textbf{4}, 1251--1258 (2017).

\bibitem{limx2019}
M.~Li, H.~Liang, R.~Luo, Y.~He, and Q.~Lin, \enquote{High-q 2d lithium niobate
  photonic crystal slab nanoresonators,} {\protect\JournalTitle{Laser \&
  Photonics Reviews}} \textbf{13}, 1800228 (2019).

\bibitem{lim2019}
M.~Li, H.~Liang, R.~Luo, Y.~He, J.~Ling, and Q.~Lin, \enquote{Photon-level
  tuning of photonic nanocavities,} {\protect\JournalTitle{Optica}} \textbf{6},
  860--863 (2019).

\bibitem{saravi2015}
S.~Saravi, S.~Diziain, M.~Zilk, F.~Setzpfandt, and T.~Pertsch,
  \enquote{Phase-matched second-harmonic generation in slow-light photonic
  crystal waveguides,} {\protect\JournalTitle{Physical Review A}} \textbf{92},
  063821 (2015).

\bibitem{Yoda2020}
T.~Yoda and M.~Notomi, \enquote{Generation and annihilation of topologically
  protected bound states in the continuum and circularly polarized states by
  symmetry breaking,} {\protect\JournalTitle{Physical Review Letters}}
  \textbf{125}, 053902 (2020).

\bibitem{lari2021}
T.~Yoda and M.~Notomi, \enquote{Generation and annihilation of topologically
  protected bound states in the continuum and circularly polarized states by
  symmetry breaking,} {\protect\JournalTitle{Physical Review Letters}}
  \textbf{125}, 053902 (2020).

\bibitem{chenya2021}
Y.~Chen, Z.~Lan, J.~Li, and J.~Zhu, \enquote{Topologically protected second
  harmonic generation via doubly resonant high-order photonic modes,}
  {\protect\JournalTitle{Physical Review B}} \textbf{104}, 155421 (2021).

\end{thebibliography}





\section*{Appendix}
\begin{figure}[!ht]
\centering
\includegraphics[width=0.55\linewidth]{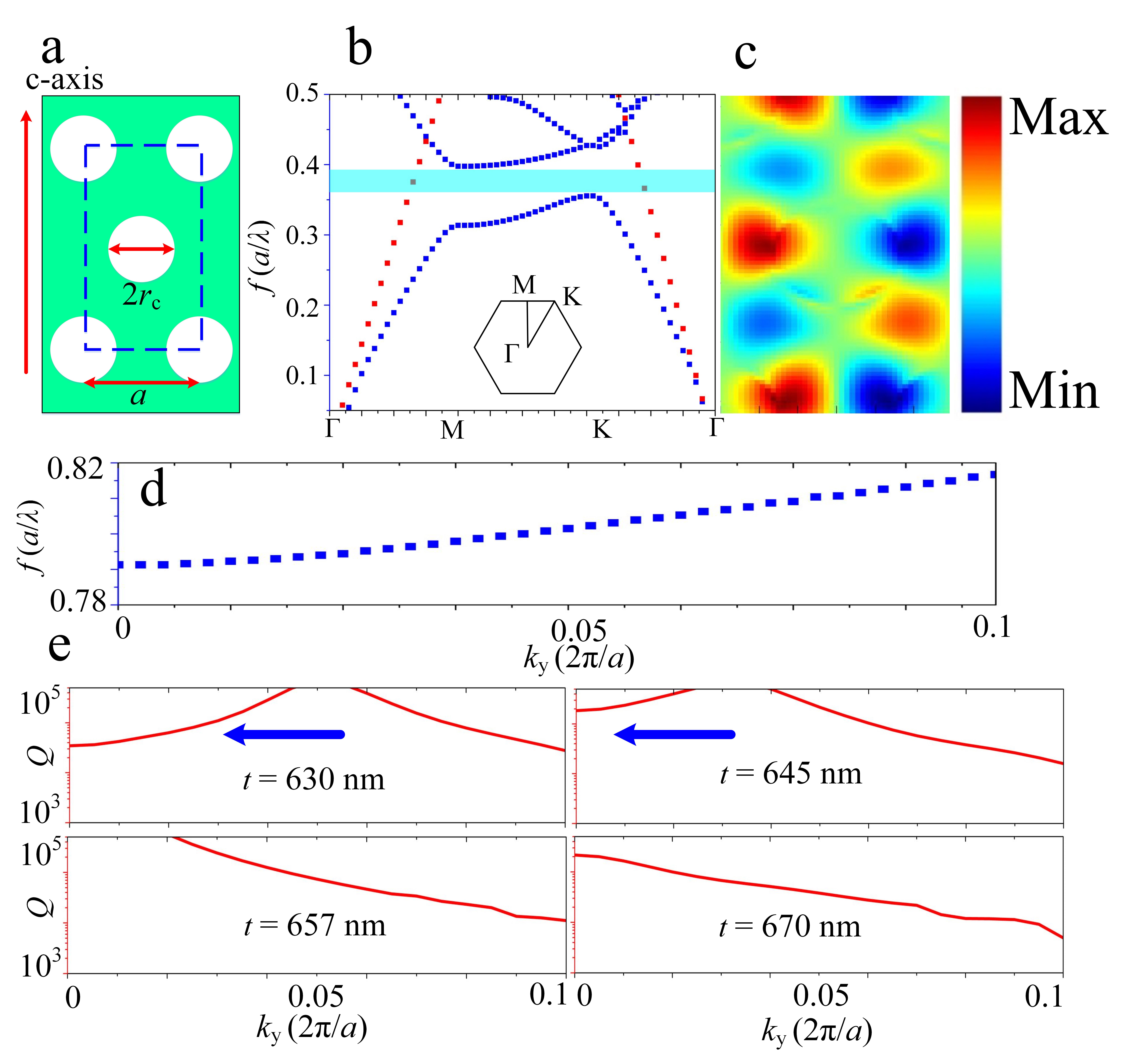}
\caption{(a) Model of PhC with optical axis pointing to $y$ direction. (b) Band diagram at the fundamental frequency. (c) Field profiles of the merged BICs mode at the second-harmonic frequency. (d) Band diagram at the second-harmonic frequency. Evolution of the accidental BIC when the thickness is (e) 630 nm, (f) 645 nm, (g) 657 nm, and (h) 670 nm.}
\label{figapp}
\end{figure}

As we discussed in the main text once $d_{33}$ tensor is used the conversion efficiency is higher. Here we consider the structure in Fig. \ref{figapp}(a) with the optical axis pointing to $y$ direction. We set the lattice constant $a$=630 nm, the thickness $t$=630 nm, and the radius of holes $r$=170 nm. We mainly consider the TE band gap in both fundamental and second-harmonic frequencies. Moreover, the band diagram at the fundamental frequency is shown in Fig. \ref{figapp}(b) and the band gap is labeled in the blue box. The $H_{z}$ field of the BIC mode is plotted in Fig. \ref{figapp}(c) and the band diagram at second-harmonic frequency is shown in Fig. \ref{figapp}(d). Near 2$f_{band-edge}$ we can find an accidental BIC and by changing the thickness of the slab the accidental BIC is approaching $\Gamma$ point as shown in Fig. \ref{figapp}(e). When the thickness is 657 nm the merged BICs can be obtained and no symmetry-protected BIC can be found. The field of the merged BICs mode shows the odd symmetry along the $y$ axis, and consequently, the nonlinear overlapping factor using $d_{33}$ is zero \cite{minkov2019}. We sweep numerous geometry parameters of PhC and find that it is hard to obtain the accidental BIC which satisfies this condition.

\end{document}